\documentclass{aa}%[referee]

\usepackage{natbib}
\usepackage{graphicx}
\usepackage[mathcal]{eucal}
\usepackage{amssymb} 
\usepackage{revsymb}	
\usepackage{amsmath}
\usepackage[usenames]{color}	
\usepackage{epstopdf}	
\newcommand{\ds}{\displaystyle}

\long\def\jumpover#1{{}}

\newcommand{\eq}[1] {Eq.~(\ref{#1})}

\newcommand{\fig}[3]{
      \begin{figure}[ht]
        \begin{center}
        \resizebox{\hsize}{!}{\includegraphics  {#1}}
        \end{center}    
        \caption{#2}
        \label{#3}
        \end{figure} }
\newcommand{\eqn} [1] {
\begin{equation}#1
\end{equation}}
\newcommand{\eqna} [1] {
\begin{eqnarray}#1
\end{eqnarray}}

\usepackage[applemac]{inputenc}

\def\acenA{$\alpha$~Cen~A}

\begin{document}

\title{The CoRoT\thanks{The CoRoT space mission, launched on
    December 27 2006, has been developped and is operated by CNES,
    with the contribution of Austria, Belgium, Brasil, ESA, Germany
    and Spain.} target HD~49933:} 
\subtitle{1 - Effect of the metal abundance on the mode excitation rates}

\author{
 R. Samadi \inst{1}   
\and H.-G. Ludwig\inst{2}
\and K. Belkacem\inst{1,3}  
\and M.J. Goupil\inst{1}    
\and M.-A. Dupret\inst{1,3}  
}
\institute{
Observatoire de Paris, LESIA, CNRS UMR 8109, Universit\'e Pierre et Marie Curie, Universit\'e Denis Diderot, 5 pl. J. Janssen, F-92195 Meudon, France  \and 
Observatoire de Paris, GEPI, CNRS UMR 8111, 5 pl. J. Janssen, F-92195 Meudon, France \and 
Institut d'Astrophysique et de Géophysique de l'Université de Liège,
Allée du 6 Août 17 – B 4000 Liège, Belgium
}

%\offprints{Reza.Samadi@obspm.fr}
\mail{Reza.Samadi@obspm.fr}

\date{\today} % Received / Accepted}

\titlerunning{The CoRoT target HD~49933: 1 - Role of the metal abundance}

%%%%%%%%%%%%%%%%%%%%%%%%%%%%%%%%%%%%%%%%%%
\abstract
  % context heading (optional)
{Solar-like oscillations are stochastically  excited by turbulent convection at the
  surface layers of the stars.}
  % aims heading (mandatory)
{We study the role of the surface metal abundance on the
  efficiency of the stochastic driving in the case of the CoRoT target HD~49933.}
  % methods heading (mandatory)
{We compute  two 3D hydrodynamical simulations representative --~ in
effective temperature and gravity ~-- of the
surface layers of the CoRoT target HD~49933, a star that is rather metal poor and
significantly hotter than the Sun. One 3D simulation has a solar metal abundance, and
the other  has a surface iron-to-hydrogen, [Fe/H], abundance ten times
smaller. For each 3D simulation we match an associated global 1D
model, and we compute the  associated acoustic modes using a theoretical
model of stochastic excitation validated in the case of the Sun and {\acenA}.}
  % results heading (mandatory)
{The rate at which energy is supplied per unit time into the acoustic modes
associated with the 3D simulation with [Fe/H]=-1 is found to be about three times smaller than those
associated with the 3D simulation with [Fe/H]=0.
As shown here, these differences are related to the fact that low metallicity
implies surface layers with a higher mean density. In turn, a higher
mean density favors smaller convective velocities and hence  less
efficient driving of the acoustic modes. }
  % conclusions heading (optional),
{Our result shows the importance of taking the surface
metal abundance into account in the modeling of the mode driving by
turbulent convection.
A comparison with observational data is presented in a companion paper using seismic data obtained for the CoRoT target HD~49933. }  


\maketitle

%%%%%%%%%%%%%%%%%%%%%%%%%%%%%%%%%%%%%%%%%%
\section{Introduction}

Using the measured linewidths and the amplitudes of the solar acoustic modes,
it has been possible  to infer the rate at which energy 
is supplied per unit time  into the solar acoustic modes. Using these constraints, 
 different models of mode excitation by turbulent
convection have been extensively tested in the case of the Sun (see
e.g. recent reviews by \citet{Samadi08b}  and \citet{Houdek06}). 
Among the different approaches, we can distinguish pure theoretical
approaches \citep[e.g.][]{Samadi00I,Chaplin05}, semi-analytical approaches  \citep[e.g.][]{Samadi02I,Samadi02II} and pure numerical approaches  \citep[e.g.][]{Stein01I,Stein04,Jacoutot08a}. 
The advantage of a theoretical approach  is that it easily allows 
 massive computation of the mode excitation rates for a wide variety
of stars with different  fundamental parameters (e.g. effective
temperature, gravity) and different surface metal abundance. 
However, pure  theoretical approaches are based on crude or simplified descriptions
of turbulent convection.
On the other hand, a semi-analytical approach is generally more
realistic since the quantities related to turbulent convection 
 are obtained from 3D hydrodynamical simulation. 
3D hydrodynamical simulations are at this point in time too time consuming, so
that a fine grid of 3D models with a sufficient resolution in effective temperature ($T_{\rm
  eff}$), gravity ($\log g$) and surface metal abundance ($Z$) is not yet available. 
In the present paper, we study and provide a procedure to interpolate
for any value of $Z$ the mode excitation
rates ${\cal P}$ between two 3D simulations with
different $Z$ but the same $T_{\rm   eff}$ and $\log g$.
With such interpolation procedure it is no longer required to have at
our disposal a fine grid in $Z$ of 3D simulations.

The semi-analytical mode that we consider here is based on  \citet{Samadi00I}'s
theoretical model with the improvements proposed by \citet{Kevin06a}. This semi-analytical model satisfactorily reproduces the solar seismic data 
\citep[][]{Samadi02II,Kevin06b}. 
Recently, the seismic constraints obtained for {\acenA} (HD 128620) have
provided an additional validation of the basic  physical
assumptions of this theoretical model \citep{Samadi08}. 
The star {\acenA} has a surface gravity ($\log g
=4.305$) lower than that of the Sun ($\log~g_\odot =$~4.438), but its
effective temperature ($T_{\rm eff}=5810$~K)  does not 
significantly differ from that of the Sun ($T_{\rm   eff,\odot}=5780$~K). 
 The higher $T_{\rm eff}$, the more vigorous the convective
velocity at the surface and the stronger the driving by turbulent
convection \citep[see e.g.][]{Houdek99}. 
For main sequence stars with a mass $M \lesssim 1.6 \, M_\odot$, an increase of the
convective velocity is expected to be associated with a larger  turbulent
Mach number, $M_t$ \citep{Houdek99}.  However, the  theoretical models of stochastic excitation  are
strictly valid in a medium where $M_t$ is --~ as
in the Sun and \acenA ~--  rather small. Hence, the higher $M_t$, the more
questionable the different approximations and the assumptions involved in the
theory \citep[see e.g.][]{Samadi00I}. 
It is therefore important to test the theory with another star characterized by a $T_{\rm   eff}$ significantly higher
  than in the Sun.
 
Furthermore, the star {\acenA} has an iron-to-hydrogen abundance slightly
larger than the Sun, namely [Fe/H]= 0.2 \citep[see][]{Neuforge-Verheecke97}. However, the modeling performed by
\citet{Samadi08} for {\acenA} assumes a solar iron abundance ([Fe/H]=0).  
According to \citet{Houdek99}, the mode amplitudes are expected to
change with the metal abundance. 
However, \citet{Houdek99}'s result was obtained on the basis of a mixing-length
approach involving several free parameters and by using a theoretical model of
stochastic excitation  in which a free multiplicative factor is introduced in
order to reproduce the maximum of the  solar mode excitation rates.  
Therefore, it is important to extend \citet{Houdek99}'s study by using
a more realistic modeling based on  3D hydrodynamical simulation of
the surface layers of stars and a theoretical model of mode driving
that reproduces ~-- without the introduction of free parameters --~ the
available seismic constraints.

To this end, the star  HD~49933 is an interesting
case for three reasons:
First, this star has $T_{\rm eff} = 6780 \pm 130$
K \citep{Bruntt08},  $\log g \simeq 4.25 \pm 0.13$ \citep{Bruntt08}
and  [Fe/H] $\simeq -0.37$ dex
\citep{Solano05,Gillon06}. 
The properties of its surface layers are thus significantly different from
those of the Sun and $\alpha$~Cen~A. 
Second, HD~49933 was observed in Doppler velocity  with the 
HARPS spectrograph. A seismic analysis of these data performed by
\citet{Mosser05} has provided the maximum of the mode surface velocity
($V_{\rm max}$). 
Third, the star was more recently  observed continuously in intensity by CoRoT during 62 days. 
Apart from observations for the
Sun, this is the longest seismic observation ever peformed both from
the ground and from space.
This long term and \emph{continuous} observation provides a very high
frequency resolution ($\sim 0.19\,\mu$Hz). The seismic analysis of these
observations undertaken by \citet{Appourchaux08} or more recently by \citet{Benomar09b}  have provided the
\emph{direct}  measurements of the mode amplitudes and the 
mode linewidths  with an accuracy
not previously achieved for a  star other than the Sun.

We consider  two 3D hydrodynamical simulations 
representative --~ in effective temperature and gravity ~-- of the
surface layers of HD~49933. One 3D simulation has  [Fe/H]=0, while the
second has [Fe/H]= -1. For each 3D simulation, we match an associated global 1D
model and  compute the  associated acoustic modes and mode excitation
rates, ${\cal P}$.  This permits us to quantify  the variation of
${\cal P}$ induced by a change of the surface metal abundance $Z$. 
From these two sets of calculation, we then deduce ${\cal P}$ for
HD~49933 by taking into account the observed iron abundance of the 
 star (i.e. [Fe/H]=-0.37). 
In a companion paper \citep[][hereafter Paper~II]{Samadi09b}, we will use these theoretical calculations
of ${\cal P}$ and the mode linewidths obtained from
 the seismic analysis of HD 49933 performed with the CoRoT data to derive the expected
 mode amplitudes in HD 49933. 
These computed mode amplitudes will then be compared with the observed
ones. This comparison will then constitute a test of the stochastic
excitation model  with a star significantly different from the Sun and
{\acenA}. It will also constitute a test of the procedure proposed
here for deriving  ${\cal P}$ for any value of $Z$ between two 3D simulations with
different $Z$.

The present paper is organised as follows:
we first describe in Sect.~\ref{calculation_P} the method  to compute
the theoretical mode excitation rates associated with the
two 3D hydrodynamical simulations.
Next, the effects on ${\cal P}$ of a different surface metal
abundance  are presented in Sect.~\ref{metal_effect}.
Then, by taking into account the actual iron abundance of  
 HD~49933, we derive theoretical values of ${\cal P}$ expected for
 HD~49933.  
Finally, Sect.~\ref{Conclusion} is dedicated to our conclusions.

%************************************
\section{Calculation of mode excitation rates}
\label{calculation_P}

%------------------------------------
\subsection{Model of stochastic excitation}
\label{The_model}

The energy injected into a mode per unit time ${\mathcal P}$  is
given by the relation \citep[see][]{Samadi00I,Kevin06b}:
\begin{equation}
\label{power}
{\mathcal P} = \frac{1}{8 ~ \ds{I} } \left ( C_R^2 + C_S^2  \right )
\;,
\end{equation}
where  $C_R^2$ and $C_S^2$ are the turbulent 
Reynolds stress and entropy contributions, respectively, and 
%$\ds{I} = \int_0^M d m \, \| \xi_r \| ^2 $
\eqn{I = \int_0^M {\rm d  m} \, \vert \xi_r \vert^2  
\label{inertia}
}
is the mode inertia, $\xi_r$ is the adiabatic radial mode displacement and $M$ is the mass of
the star.
The expressions for $C_R^2$ and $C_S^2$ are given  for a radial mode with
frequency $\omega_{\rm osc}$ by
\begin{eqnarray}
\label{C2R}
C_R^2 & =  & { 64 \pi^{3} \over 15 } \int {\rm d m}  \, {  {\bar \rho} \, {\tilde u}^4  \over
  {k_0^3  \, \omega_0} }   \,   {{\cal K}_w  \over 3} \, f_r \, S_R(r,\omega_{\rm osc}) \; ,\\
\label{C2S}
C_S^2 &=& \frac{16 \pi^3}{3 \, \omega_{\rm osc}^2}  \int {\rm d m}  \, {
    { (\alpha_s \, \tilde s \, {\tilde u})^2 }\over { {\bar \rho} \, k_0^3 \,
      \omega_0 }  } \,  g_r  \, \, S_s(r,\omega_{\rm osc})
\end{eqnarray}
where we have defined the ``source functions'':
\begin{eqnarray}
\label{SR}
S_R(r,\omega_{\rm osc}) &=& { { k_0^3 \, \omega_0 } \over {\tilde u}^4}  \,\int
\frac { {\rm d} k } {k^2 }\,  E^2(k)  \nonumber \\ &  & \;  \times\int
{\rm d} \omega \,  \chi_k( \omega + \omega_{\rm osc}) ~\chi_k( \omega )
 \\
S_s(r,\omega_{\rm osc}) & = & { { k_0^3 \,\omega_0} \over {{\tilde
      u}^2 \, \tilde{s}^2}} \, \int \frac{ {\rm d} k}{k^2}\,  E(k) 
\,  E_s(k)  \nonumber \\ &  & \;  \times\int
{\rm d} \omega \, \chi_k(\omega+\omega_{\rm osc})\,  \chi_k(\omega)
\label{SS}
\end{eqnarray}
where  $P$ is the gas pressure, $\rho$
the density, $s$ the entropy,  ${\bar \rho}$ the equilibrium density
profile, $\alpha_s \equiv (\partial P / \partial s)_\rho$,  $f_r
\equiv ({\rm d} \xi_r / {\rm d} r)^2$ and $g_r$ are two functions
 that involve the first and second derivatives of
$\xi_r$ respectively,
 $k$  is the wavenumber,  ${E}(k)$ is
the turbulent kinetic energy spectrum, ${E}_s(k)$ is
the spectrum
associated with the entropy fluctuations ($s$), $\tilde s$ is the rms
of $s$, $\chi_k$ is the time-correlation function associated with the velocity,  ${\tilde u}$ is a characteristic velocity 
defined in a way that $3 \,  {\tilde u}^2 =  \langle \vec{u}^2\rangle  $,  $\langle.\rangle$ refers to 
horizontal and time average, $\vec{u}$ is the turbulent velocity 
field, and finally ${\cal K}_w\equiv \langle u_z^4\rangle/\langle u_z^2\rangle^{2}$ is the Kurtosis 
\citep[see][ for details]{Kevin06a,Kevin06b}. 
Furthermore,  we have introduced for convenience the
characteristic frequency $\omega_0$  and  the characteristic wavenumber
$k_0$: % ; they are defined as:
\eqna{
 \omega_0 & \equiv & k_0 \, {\tilde u} \label{omega_0}  \label{u0}\\
 k_0 &\equiv &{ {2 \pi} \over \Lambda} \label{k_0}   \label{k0}
}
where  $\Lambda$ is a characteristic size derived from $E(k)$ as
explained in \citet{Samadi02I}.
Note that the introduction of the term ${\ds  k_0^{3}  \, \omega_0 \,
   \, {\tilde u}^{-4}} $  in the RHS of Eq.~(\ref{SR}) and the 
term ${ \ds  k_0^3 \,\omega_0 \,  {\tilde u}^{-2} \, \tilde{s}^{-2}}$ in the
RHS of Eq.~(\ref{SS}) ensure dimensionless source functions.

% Accordingly, we have ${\tilde u}
%= \sqrt{\Phi/3} \, w$. For isotropic convection, the  characteritic velocity
%${\tilde u} w$  is such that  $\langleu_x^2\rangle= \langleu_y^2\rangle= \langleu_z^2\rangle= w^2 = {\tilde u}^2 $.

The kinetic spectrum $E(k)$ is derived from the 3D simulation as detailled in  \citet{Samadi02I}.
As shown by \citet{Samadi02I}, the $k$-dependence of $E_s(k)$ is
similar to that of the $E(k)$. Accordingly, we assume $E_s \propto
E$.

In \citet{Samadi08}, two different analytical functions for
$\chi_k(\omega)$ have been considered, namely a 
Lorentzian function and a Gaussian one. In the present study we will
in addition derive $\chi_k(\omega)$ \emph{directly} from the 3D simulations
as detailled in \citet{Samadi02II}. Once $\chi_k(\omega)$ is derived
from the 3D simulation, it is implemented in Eq.~(\ref{SR}) and
Eq.~(\ref{SS}).
% Note that this requires to compute the auto-correlation of
%$\chi_k(\omega)$.

We compute the mode excitation as detailled in
\citet{Samadi08}: all required quantities --~ except $\xi_r$, $I$ and
$\omega_{\rm osc}$ ~--  are   obtained  \emph{directly}  from two 3D hydrodynamical simulations
 representative of the outer layers of HD~49933, whose characteristics are described in
Sect.~\ref{3Dmodels} below.

The quantities related to the modes ($\omega_{\rm osc}$, $I$ and $\xi_{\rm
  r}$)  are calculated using  the adiabatic pulsation code ADIPLS
\citep{JCD91b} from  1D global models.
The outer layers of these 1D models are derived from the 3D
simulation as described in Sect.~\ref{1Dmodels}.

%------------------------------------
\subsection{The 3D simulations}
\label{3Dmodels}

We computed two 3D radiation-hydrodynamical model atmospheres with the code
CO$^5$BOLD \citep{Freytag02,Wedemeyer04}. One 3D simulation had a solar
iron-to-hydrogen [Fe/H]=0.0 while the other had [Fe/H]=-1.0. The 3D model with
[Fe/H]=0 (resp. [Fe/H]=-1) will be hereafter referred to as model S0 (resp. S1). The
assumed chemical composition is similar (in particular for the CNO elements)
to that of the solar chemical composition proposed by \citet{Asplund05}. The
abundances of the $\alpha$-elements in model~S1 were assumed to be enhanced by
0.4\,dex. For S0 we obtain $Z/X$ = 0.01830 and Y=0.249, and for S1 $Z/X$ =
0.0036765 and $Y$=0.252.  Both 3D simulations have exactly the same gravity
($\log g=$4.25) and are very close in effective temperature ($T_{\rm eff}$).
Both models employ a spatial mesh with $140 \times 140 \times 150$ grid
points, and a physical extent of the computational box of $16.4 \times 16.4 \times
24.2$~Mm$^3$.  The equation of state takes into account the ionisation of
hydrogen and helium as well as the formation of H$_2$ molecules according to
the Saha-Boltzmann statistics. The wavelength dependence of the radiative transfer
is treated by the opacity binning method \citep{Nordlund82,Ludwig92,Voegler04}
using five wavelength bins for model S0 and six for model S1. Detailed wavelength-dependent opacities were obtained from the MARCS model atmosphere package   
\citep{Gustafsson08}.  Table~\ref{tab:3Dmodels} summarizes the
characteristics of the 3D models.  The effective temperature and surface
gravity correspond to the parameters of HD~49933 within the observational
uncertainties, while the two metallicities bracket the observed value.

For each 3D simulation, two time series were built. One has a long duration
(38h and 20h for S0 and S1, respectively) and a low sampling
frequency (10~mn). This time series is used to compute time averaged
quantities (${\bar \rho}$, $E(k)$, etc.).  The second time series is shorter
(8.8h and 6.8h for S0 and S1, respectively), but has a high
sampling frequency (1~mn). Such high sampling frequency is required
for the calculation of $\chi_k(\omega)$. Indeed, the modes we are looking at lie between $\nu
\approx 1.25$~mHz and $\nu \approx 2.4$~mHz. 

\begin{table}[ht]
\begin{center}
\begin{tabular}{ccccccc}
 Label & [Fe/H]   & $Y$ & $Z$ &  $Z/X$ & $T_{\rm eff}$ [K]  \\
\hline
 S0 & 0 &   0.249 &  $13.5~10^{-3}$ & 0.018305  & 6725 $\pm$ 17\\
 S1 & -1 &   0.252 &  $2.74~10^{-3}$ & 0.003676 & 6730 $\pm$ 12 
\end{tabular}
\end{center}
\caption{Characteristics of the 3D simulations. }
\label{tab:3Dmodels}
\end{table}

The two 3D simulations extend up to $T=100\,000~$K. However, for $T
\gtrsim 30\,000~$K, the 3D simulations are not completely realistic. First of
all, the MARCS-based opacities are provided only up to a temperature of
30\,000~K; for higher temperatures the value at 30\,000~K is assumed. Note
that we refer to the opacity per unit mass  here. For the radiative transfer
the opacity per unit volume is the relevant quantity, i.e. the product of
opacity per mass unit and density. Since in the simulation the opacity is
still multiplied at each position with the correct local density, the actual
error we make when extrapolating the opacity is acceptable. 

Another limitation of the simulations is the restricted size of the computational box which does
not allow for a full development of the largest flow structures, again in the
layers above $T \simeq 30~000$~K. Two hints make us believe that the size
of the computational domain is not fully sufficient: i) in the deepest layers
of the simulations there is a tendency that structures align with the
computational grid; ii) the spatial spectral power~$P$ of scalar fields in a horizontal layer 
does not tend towards the expected asymptotic  behaviour $P \times k$ for low 
spatial wavenumber~$k$.  We noticed this shortcoming only after the completion of the simulation
runs. 
To mitigate its effect in our analysis, we will later by default integrate the
mode excitation rates up to $T = 30\,000$~K. However, for comparison purposes,
some computations have been extended down to the bottom of the 3D
simulations. For S0, the layers 
 located below $T \simeq 30~000$~K contribute only by  $\lesssim 10$\,\% to the excitation of the 
modes lying in the frequency range where modes have the most chance to
be detected ($\nu \simeq 1.2 - 2.5$ mHz). For S1, the contribution of the deep layers
is even smaller ($\sim 5$\,\%). 

%Note that in both 3D simulations, the turbulent Mach number, $M_t$,
%reaches the maximum value of $\approx$ 0.45; for comparison it is up
%to $\approx 0.32$ for equivalent solar simulations.

Finally, one may wonder how the treatment of the small-scales or the
limited spatial resolution of the simulation can influence our
calculations. Dissipative processes are handled in CO$^5$BOLD   on  the one hand 
side implicitely by the numerical scheme (Roe-type approximate Riemann 
solver), and on the other hand explicitely by a sub-grid model according to 
the classical \citet{Smagorinsky63} formulation. 
\citet{Jacoutot08a} found that computed mode
excitation rates significantly depend on the adopted sub-grid model. 
%However,  they obtained this influence in simulation of quite low resolution ($66 
%\times 66 \times 40$ grid cells) where the relative influence of the sub-grid 
%scale model is particularly pronounced. 
\citet{Samadi07a} have found that solar mode excitation rates computed in the manner of \citet{Stein01I},
i.e., using  data directly  from the  3D simulation,
decrease as the spatial resolution of the solar 3D simulation decreases.
As a conclusion the spatial resolution or the sub-grid
model can influence computed mode excitation rates \citep[see a discussion
in][]{Samadi08}. 
However, concerning the spatial resolution and according to
\citet{Samadi07a}'s results, the present spatial resolution (1/140 of
the horizontal size of the box and about 1/150 of the vertical extent
of the simulation box) is high enough to obtain  accurate computed  energy 
rates. The increased spatial resolution of our models in comparison to the 
work of  \citet{Jacoutot08a}  reduces the impact of the unresolved scales.

%------------------------------------
\subsection{The 1D global models}
\label{1Dmodels}

For each 3D model we compute an associated 1D global model.
The models  are built in the manner of \citet{Trampedach97} as detailled in \citet{Samadi08}
in such way that their outer layers are replaced by the averaged 3D
simulations described in Sect.~\ref{3Dmodels}. 
The interior of the models are obtained with the CESAM code
assuming standard physics: Convection  is 
described according  to  \citet{Bohm58}'s   local mixing-length theory of convection (MLT), 
and   turbulent pressure is ignored. 
Microscopic diffusion is not included.
The OPAL equation of state is assumed. The chemical
mixture of the heavy elements is similar to that  of \citet{Asplund05}'s mixture. 
As in  \citet{Samadi08}, we will refer to these models as ``patched''
models hereafter.

 The two models have  the effective temperature and the gravity of the
 3D simulations. One model is matched with S0 and has [Fe/H]=0, while the
 second is matched with S1 and has [Fe/H]=-1. The 1D models have the
 same chemical mixture as their  associated 3D simulations.
The parameters of the 1D patched models are given in  Table~\ref{tab:1Dmodels}.
 The  stratification in density and 
temperature of the patched 1D models are shown in Fig.~(\ref{rho0}). 
At any given temperature the density is larger in S1  as a consequence of
its lower metal abundance.
Indeed, the lower the metal abundance, the lower the opacity ; then,
at a given optical depth ($\tau$), the density is larger in S1 compared to S0.
The photosphere corresponds to the optical depth $\tau = 2/3$. Since
the two 3D simulations have approximatively the  same effective
temperature, the density in S1 is larger at optical depth $\tau =
2/3$. Since the density in S1 increases with depth even more
rapidly than in S0, the density in S1 remains larger for   $\tau >
2/3$ than in S0.
\begin{table*}[ht]
\begin{center}
\begin{tabular}{cccccccc}
 [Fe/H]   & $Y$ & $Z$ &  $T_{\rm eff}$ [K]  &$R/R_\odot$ & $M/M_\odot$ & $\alpha$  \\
\hline
  0 &   0.249 &  $13.5~10^{-3}$ & 6726 & 1.473 &  1.408 & 1.677 \\
 -1 &   0.252 &  $2.74~10^{-3}$ & 6732 & 1.261 &  1.033 & 1.905 
\end{tabular}
\end{center}
\caption{Characteristics of the 1D ``patched'' models.  $\alpha$ is the
mixing-length parameter. }
\label{tab:1Dmodels}
\end{table*}

\fig{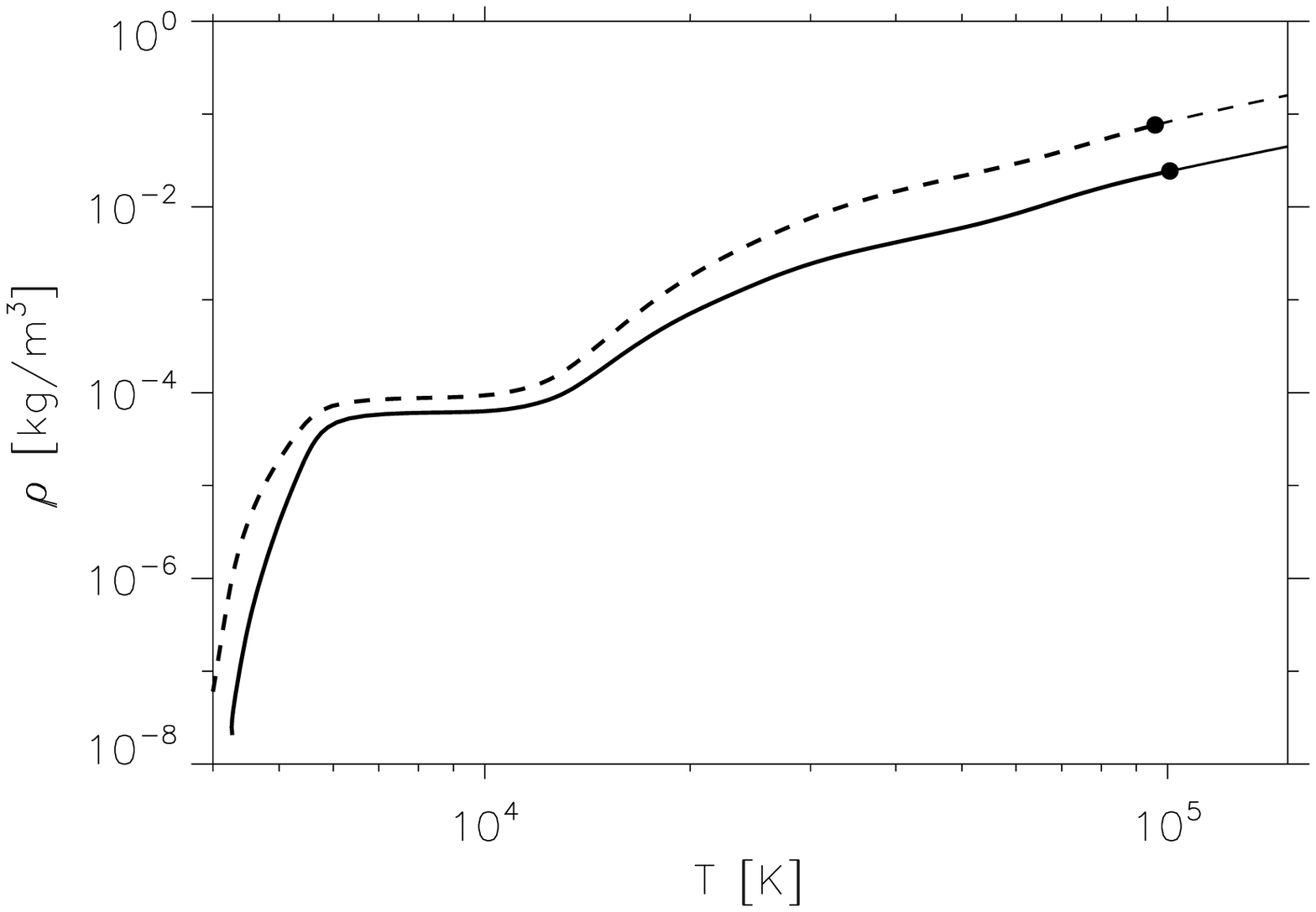}{Mean density ${\bar \rho}$  as a function of
  temperature, $T$. The solid line corresponds to the 3D model
with the metal abundance (S0) and the dashed line to metal poor 3D
model (S1).  The filled dots show the location
  where the 1D models have been matched to the associated 3D simulation.}{rho0}

%************************************
\section{Effects of the metal abundance on excitation rates}

\label{metal_effect}

The mode excitation rates (${\cal P}$) are computed for the two 3D simulations according
to Eqs.~(\ref{power})-(\ref{SS}). The integration is performed from
the top of the simulated domains down to $T = 30\,000$~K (see
Sect.~\ref{3Dmodels}). 
 In the following, ${\cal P}_1$ (resp. ${\cal P}_0$) corresponds to the mode
excitation rates associated with the 3D model with [Fe/H]=-1  (resp. [Fe/H]=0)
%...........................
\subsection{Results}

\label{metal_effect_results}

Figure~\ref{pow} shows the effect of the assumed metal abundance of the
stellar model on the mode
excitation rates. ${\cal P}_1$ is found to be three times smaller than
${\cal P}_0$, i.e.  p modes associated with  the metal
poor 3D model (S1) receive approximatively three times less
energy per unit time than those associated with the 3D model with
the solar  metal abundance (S0).

For both 3D models, the dominant part of the
driving is ensured by the Reynolds stresses. The entropy fluctuations
contribute by only $\sim$~30~\% of the total power for both S0 and S1. By comparison,
in the case of the Sun and $\alpha$~Cen~A it contributes by only  $\sim$~15~\%.
Furthermore,  we find that the contribution of the entropy source term
is --~ as for the Reynolds stress term ~-- about three times smaller in S1
than in S0.
We conclude that the effect of the  metal abundance
on the excitation rates is almost the same for the Reynolds stress
contribution and the entropy source term.

\fig{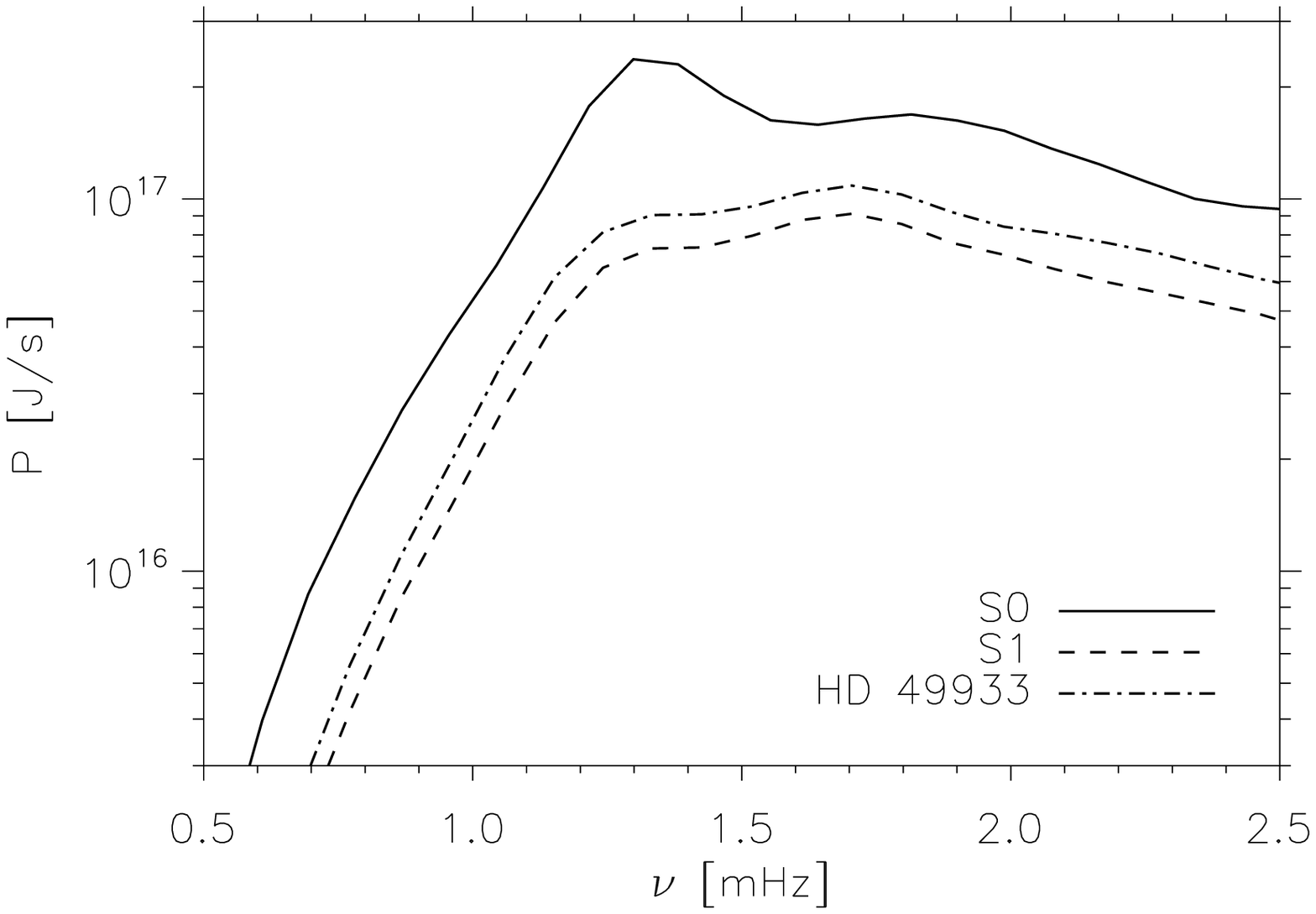}{Mode excitation rates $\mathcal P$ as a function
of the mode frequency, $\nu$. The solid line corresponds to the 3D model
with the canonical metal abundance (S0) and the dashed line to the metal poor 3D
model (S1). The dot-dashed  line corresponds to the mode excitation
rates derived for the specific case of HD~49933 as explained
in Appendix~A. 
}{pow}

%...........................
\subsection{Interpretation}
\label{metal_effect_interpretation}

From Eqs.~(\ref{power}),  (\ref{inertia}), (\ref{C2R}), (\ref{omega_0})
and (\ref{k_0})  we show that
at a given layer the power supplied to the modes  --~ per unit mass
~-- by the \emph{Reynolds  stress} is  proportional  to
$F_{\rm kin} \,  \Lambda^4 \, S_R / {\cal M}$, where 
$F_{\rm kin}$ is the flux of the kinetic energy, which  is
proportional to ${\bar \rho} \, {\tilde u}^3 $, $\Lambda$ is a
characteristic length (see Sect.~\ref{The_model}) and ${\mathcal M}$ is the mode mass defined as:
\eqn{
{\mathcal M}= { I \over { \xi_r^2 } }
\label{modemass}
} 
where $\xi_r$ is the mode displacement evaluated  at the layer  in the
atmosphere where the mode is measured.

The power  supplied to the modes --~ per unit mass ~--  
 by the entropy source term is proportional  to  ${\bar \rho} \,  {\tilde u}^3 \, \Lambda^4 \, \mathcal R^2 \,  S_s $ where
$\omega_{\rm osc}$ is the mode frequency,
%   $\tau_\Lambda \sim \Lambda / {\tilde u}$
%is the characteristic eddy turn over time, 
$\mathcal R \propto  F_{\rm conv} /
F_{\rm kin}$, where $F_{\rm conv} \propto w \, \alpha_s \,  \tilde s $
is the convective flux, and finally $\tilde s$ is the rms of the entropy fluctuations
\citep[see][]{Samadi05b}. We recall that the higher $\mathcal R$, the higher the \emph{relative}
contribution of the entropy source to the excitation.
We study below the role of  ${\cal M}$,  $F_{\rm kin}$, $\Lambda$, 
$S_R$, $S_s$ and ${\cal R}$:

\bigskip
{\it Mode mass (${\cal M}$):}
The frequency domain, where modes are strongly excited,   ranges between $\nu \approx
1.2$ mHz and $\nu \approx 2.5$ mHz.  In this frequency domain, the
mode masses ${\cal M}$ associated with S0 are quite similar to those 
associated with S1 (not shown). Consequently the differences between ${\cal P}_1$ and ${\cal P}_0$
do not arise from the (small) differences in ${\cal M}$.
 
%This is verified to be true for the two 3D simulations in most part of the
%simulated domain.
\bigskip
{\it Kinetic energy flux ($F_{\rm kin}$):}
The larger $F_{\rm kin}$, the larger the driving by the Reynolds stress.
However, we find that the two 3D models have very similar $F_{\rm
  kin}$. This is not surprising since  the two 3D models have very
similar effective temperatures. This means that the differences between ${\cal P}_1$ and ${\cal P}_0$
do not arise from the (small) differences in $F_{\rm kin}$.

\bigskip
{\it Characteristic length ($\Lambda$):}
In the manner of \citet{Samadi02I} we derive from the
kinetic energy spectra $E(k)$ of the two 3D
simulations the characteristic length $\Lambda$ ($\Lambda = 2
\pi/k_0$, see \eq{k0}) for each layer of the simulated domain. 
We find that the differences in
$\Lambda$  between the two 3D simulations is small and does not play a
significant role in the differences in $\cal P$. 
This can be understood by the fact that S0 and S1 have the same
gravity. Indeed, as shown by \citet{Samadi08} --~ at a fixed
effective temperature ~-- $\Lambda$ scales as the inverse of $g$.
We conclude that the differences between ${\cal P}_1$ and ${\cal P}_0$
do not originate from the (small) differences in  $\Lambda$. 

\bigskip
{\it Source functions ($S_R$ and $S_s$):}
The dimensionless source functions $S_R$ and $S_s$  are 
defined in Eqs.~(\ref{SR}) and  (\ref{SS}) respectively. Both source
functions involve the eddy time-correlation function
$\chi_k(\omega)$. We define  $\omega_k$  as the frequency width of
$\chi_k(\omega)$. 
As shown by \citet{Samadi02II} and as verified in
the present case, $\omega_k$ can be evaluated as the product $k \,
u_k$ where $u_k$ is given by the relation \citep{Stein67}:
\eqn{
u_k^2 = \int_k^{2k} {\rm d} k \, E(k)
\label{u_k}
}
where $E(k)$ is normalised as:
\eqna{
 \int_0^{+\infty} {\rm d } k \, E(k)   & =  {1 \over 2  }  \langle u^2\rangle  \equiv {1 \over 2  }
      {\tilde u}^2 \; .
\label{normEk}
}
According to Eqs.~(\ref{u_k}) and (\ref{normEk}), $u_k$ is directly
proportional to ${\tilde u}$.
At a fixed $k/k_0$, we  then have   $\omega_k \propto {\tilde u}
\, k_0 = \omega_0$.

As seen above,  $\omega_0$ controls $\omega_k$, the frequency width of
$\chi_k$. Then, at fixed $\omega_{\rm osc}$, we can easly see from Eqs.~(\ref{SR}) and (\ref{SS}) that the smaller  $\omega_0$, the
smaller $S_R(\omega_{\rm osc})$ and $S_s(\omega_{\rm osc})$.
Since $\omega_0 = {\tilde u} \, k_0= 2 \pi \, {\tilde u}
/\Lambda$ and since both 3D simulations have approximately the same
$\Lambda$, smaller ${\tilde u}$ results directly in smaller $\omega_0$ and
hence in smaller source functions.

We have plotted  in Fig.~\ref{figu0} the characteristic velocity
${\tilde u}$. This quantity is found to be up to 15~\%
smaller for S1 compared with S0. In other words, the metal poor 3D
model is characterized by lower convective velocities. Consequently, the
source functions are smaller for S1 compared to S0.
Although the convective velocities differ between S0 and S1 by
only 15\,\%, the excitation rates differ by a factor $\sim$~3. 
The reason for this is that he source functions, which are non-linear functions
of ${\tilde u}$,  decrease very rapidly with ${\tilde u}$. This is
the consequence of the behavior of the eddy-time correlation
$\chi_k$. Indeed, this function varies with the ratio $\omega_{\rm
  osc}/\omega_k$ approximately as a Lorentzian function. 
This is why $\chi_k$ varies rapidly with ${\tilde u}$ (we
recall that $\omega_k \propto {\tilde u} \, k_0 $).

\bigskip

In conclusion, the differences between ${\cal P}_1$ and ${\cal P}_0$
are mainly due to differences in the characteristic velocity
${\tilde u}$. 
In turn, the low convective velocity in S1 is a consequence of the
larger density compared to S0. 
Indeed, as shown in Fig.~\ref{rho0}, the density is systematically
higher in S1. At the layer where the modes are the most excited (i.e. at $T \sim 10 000 K$), the density is $\sim$~50~\% higher.
Since the two 3D models have a similar kinetic energy flux (see
above), it follows that a larger density for S1 then implies lower
convective velocities.

\fig{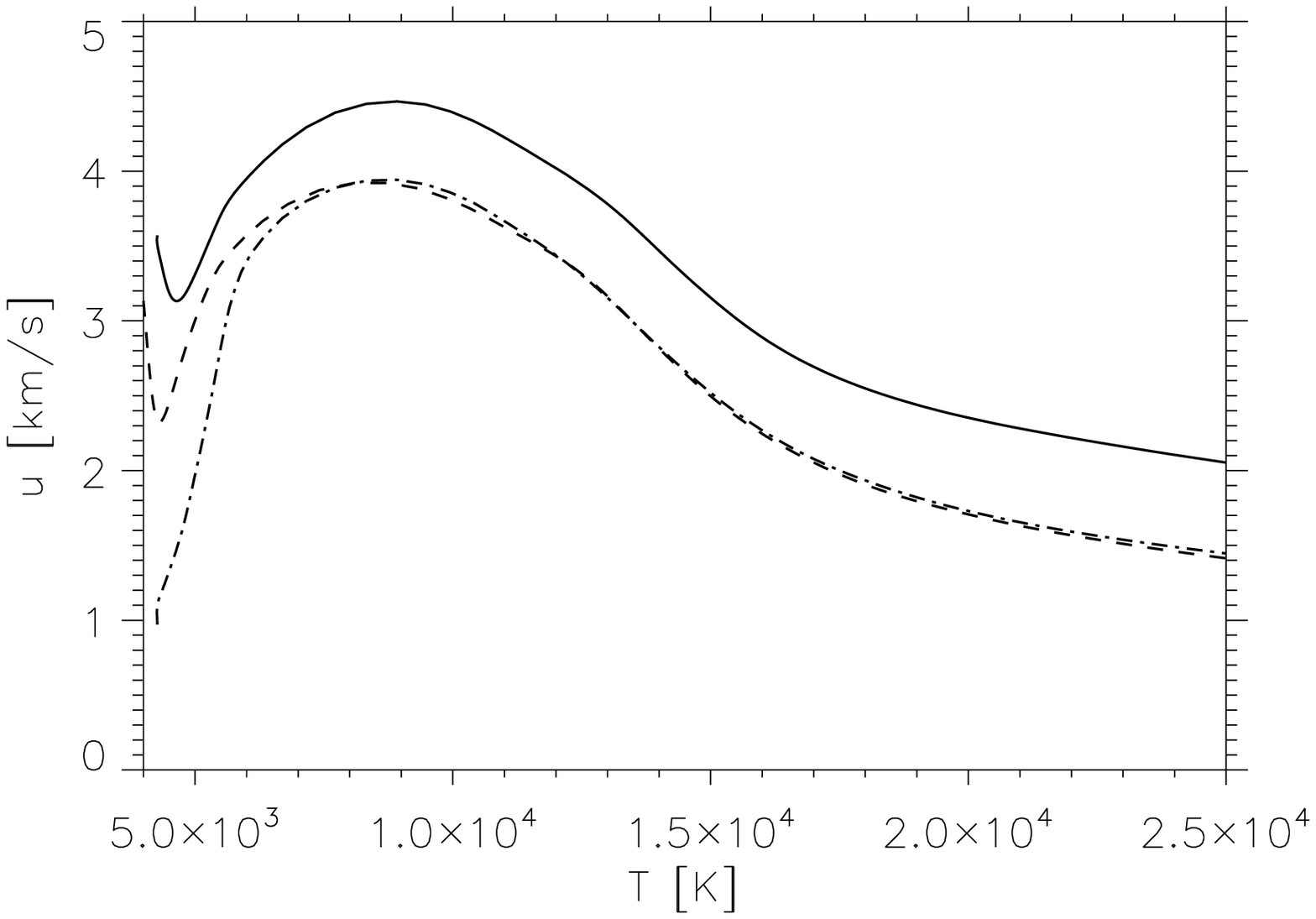}{Characteristic velocity ${\tilde u}$ defined in
  Eq.~(\ref{u0}) as a function of temperature, $T$.   The solid and dashed lines have the same
  meaning as in Fig.~\ref{pow}. The dot-dashed line corresponds to
  the solid line multiplied by $\gamma_1$, where $\gamma_1 (T) \equiv
  \left ( {\bar   \rho}_0 / {\bar \rho}_1 \right ) ^{1/3}$ and  ${\bar \rho}_0$ (resp. ${\bar
    \rho}_1$) is the mean density stratification of S0 (resp. S1)(see Appendix~A). }{figu0}

\bigskip
{\it  Relative contribution of the entropy source term (${\cal R}$):}
The convective flux $ F_{\rm conv}$ in S1 
is almost identical   to that of S0. This is due to the fact that the
two 3D simulations have almost the same effective
temperature.  Furthermore, as pointed out above, the differences in
$F_{\rm kin}$  between S1 and S0  are small.
As a consequence, the ratio $\mathcal R   \propto  F_{\rm conv} /
F_{\rm kin} $ does not differ between the two 3D simulations.
Accordingly, as for the Reynolds contribution, the variation of the excitation rates with the
metal abundance is only due to the source term $S_S$. The latter
varies with $\omega_0$ in the same manner as  $S_R$, which is turn the reason for 
As a consequence,  the contribution of the entropy fluctuations to show
the same trend with  the metal abundance as the
Reynolds stress  term.

%************************************
\section{Theoretical calculation of ${\cal P}$ for HD~49933}
\label{deriving_P}

We derive the mode excitation rates ${\cal P}$ for HD~49933. According
to \citet{Gillon06},  HD~49933 has  [Fe/H]=-0.37 $\pm$ 0.03~dex, while 
 we only have  two 3D simulations with  values of  [Fe/H], respectively
 [Fe/H]=0 and [Fe/H]=-1.

As seen in Sect.~\ref{metal_effect_interpretation}, differences in ${\cal P}$ between S0 and S1 are a
direct consequence of the differences in the source functions  $S_R$
and $S_S$. It follows that in order to derive  ${\cal P}$ for HD~49933, we only have to
derive the expected values for  $S_R$ and $S_S$.
As seen in Sect.~\ref{metal_effect_interpretation}, differences in  $S_R$ (or in 
 $S_S$) between S0 and S1 are  related to the surface metal 
abundance through the surface densities that impact the convective velocities ($\tilde u$).
 The determination of the HD 49933 convective velocities allows us to determine its source 
function. To this end, we use the fact that the kinetic flux is almost unchanged between 
S1 and S0 (see Sect.~\ref{metal_effect_interpretation}) to derive the
profile of $\tilde u (T)$, expected  at the 
surface layers of  HD~49933. This is performed by  interpolating in $Z$ between S0 and S1, 
the surface density stratification representative of the surface layers of HD~49933.
The whole procedure is described in Appendix~A.

In order to compute ${\cal P}$ for HD~49933, we then need to know
$Z$ for this star.  Since we do not know its surface helium abundance, we will
 assume  by default the solar value for $Y$:  $Y=
 0.249~\pm~$0.003 \citep{Basu97}.
\citet{Gillon06}'s analysis shows that the chemical
mixture of HD~49933 does not significantly differ from that of the
Sun.
According to \citet{Asplund05}, the new solar metal to hydrogen ratio
is $(Z/X)_\odot = 0.0165$
 Accordingly, since [Fe/H]=- 0.37 $\pm$ 0.03 dex, we derive
  $Z= 5.3~10^{-3} \pm \,0.4\,10^{-3}$  for HD~49933. 
% Using this value, we can derive ${\bar \rho}_*$, and hence $\gamma_* (T)
% \equiv ( {\bar \rho}_1 / {\bar \rho}_* )^{1/3} $. 
 Note that assuming \citet{GN93}'s chemical mixture yields $Z=
 7.8~10^{-3} \pm  ~ 0.5~10^{-3}$.    

The result of the calculation is shown in Fig.~\ref{pow}.
 The maximum ${\cal P}$ is  1.08 $\pm$ 0.05 $10^{17}$
J/s when  \citet{Asplund05}'s chemical composition is assumed (see
Appendix~A). 
This is about 30 times larger than in the Sun and about 14 times
larger than in $\alpha$~Cen~A. When \citet{GN93}'s chemical
mixture is assumed, the maximum in  ${\cal
  P}$ is in that case equal to 1.27~$\pm$ 0.05~$10^{17}$J/s, that is
about 30\,\% larger than with \citet{Asplund05}'s  solar chemical mixture. 

We  note that the uncertainties
in the knowledge of [Fe/H] set uncertainties on  ${\cal P}$ which
are the on order of 10~\% in  the frequency domain of interest.

%************************************
\section{Conclusion}
\label{Conclusion}

We have built two 3D hydrodynamical simulations representative in
effective temperature ($T_{\rm eff}$) 
and gravity ($g$) of the surface layers of an F type star  on the main
sequence. One model has a solar iron-to-hydrogen 
abundance ([Fe/H]=0) and the other has [Fe/H]=-1. Both models have the same
$T_{\rm eff}$ and $g$.
For each 3D simulation, we have
computed an associated ``patched'' 1D full model.
Finally, we have computed the mode excitation rates ${\cal P}$ associated
with the two ``patched'' 1D models.

Mode excitation rates associated with the metal poor 3D simulation are
found to be about three times smaller than those associated with the 3D
simulation which  has a solar surface metal abundance. This is explained by the following connections: the lower the metallicity, the lower the opacity. At
fixed effective temperature and surface gravity, the lower the opacity, the denser
the medium at a given optical depth. The higher the density, the smaller are the convective
velocities to transport the same amount of
energy by convection. Finally, smaller convective velocities result in a less
efficient driving.
On the other hand, a surface metal abundance higher than the solar
metal abundance will result in a lower surface density, which in
turn will result in a higher convective velocity and then in a more
efficient driving. 
Our result can then be qualitatively generalised for any surface metal
abundance.

By taking into account the observed surface metal abundance of the star HD~49933 (i.e. [Fe/H]=-0.37),
 we have derived, using two 3D simulations  and the interpolation procedure developed here, the rates at which acoustic modes
are expected to be excited by turbulent convection in the case of
HD~49933. These excitation rates ${\cal P}$ are found to be about two times
smaller than for a model built assuming a solar metal abundance.
These theoretical mode excitation rates will be used in Paper~II to derive
 the expected mode amplitudes from measured mode linewidths. We will
then be able to compare these amplitudes with those derived for HD~49933 from
different seismic data. 
 This will  constitute  an indirect test of our procedure which permits us to interpolate
for any value of $Z$ the mode excitation rates ${\cal P}$ between two 3D simulations with
different $Z$ but the same $T_{\rm   eff}$ and $\log g$. 
We must stress that a more direct validation of 
 this interpolation procedure will be to compute 
 a third 3D model with the surface metal abundance of the star HD~49933   
and to compare finally  the mode excitation rates obtained here with the interpolation procedure with that obtained with this third 3D model.
This represents a long term work since several months (about three to four months) are required for the calculation of this additional 3D model, which is in progress.

%%%%%%%%%%%%%%%%%%%%%%%%%%%%%%%%%%%%%%%%%%
\begin{acknowledgements}
 We thank C. Catala for  useful discussions
 concerning the spectrometric properties of HD~49933.
We are indebted to J. Leibacher for his careful reading of the
manuscript.   
K.B. acknowledged financial support from Li\`ege University through the Subside F\'ed\'eral pour la Recherche 2009. 
\end{acknowledgements}
\bibliographystyle{aa}
%\bibliography{/home/reza/redac/biblio.bib}

\begin{thebibliography}{36}
\expandafter\ifx\csname natexlab\endcsname\relax\def\natexlab#1{#1}\fi

\bibitem[{{Appourchaux} {et~al.}(2008){Appourchaux}, {Michel}, {Auvergne},
  {Baglin}, {Toutain}, {Baudin}, {Benomar}, {Chaplin}, {Deheuvels}, {Samadi},
  {Verner}, {Boumier}, {Garc{\'{\i}}a}, {Mosser}, {Hulot}, {Ballot}, {Barban},
  {Elsworth}, {Jim{\'e}nez-Reyes}, {Kjeldsen}, {R{\'e}gulo}, \&
  {Roxburgh}}]{Appourchaux08}
{Appourchaux}, T., {Michel}, E., {Auvergne}, M., {et~al.} 2008, \aap, 488, 705

\bibitem[{{Asplund} {et~al.}(2005){Asplund}, {Grevesse}, \&
  {Sauval}}]{Asplund05}
{Asplund}, M., {Grevesse}, N., \& {Sauval}, A.~J. 2005, in Astronomical Society
  of the Pacific Conference Series, Vol. 336, Cosmic Abundances as Records of
  Stellar Evolution and Nucleosynthesis, ed. T.~G. {Barnes}, III \& F.~N.
  {Bash}, 25

\bibitem[{{Basu}(1997)}]{Basu97}
{Basu}, S. 1997, \mnras, 288, 572

\bibitem[{{Belkacem} {et~al.}(2006{\natexlab{a}}){Belkacem}, {Samadi},
  {Goupil}, \& {Kupka}}]{Kevin06a}
{Belkacem}, K., {Samadi}, R., {Goupil}, M.~J., \& {Kupka}, F.
  2006{\natexlab{a}}, \aap, 460, 173

\bibitem[{{Belkacem} {et~al.}(2006{\natexlab{b}}){Belkacem}, {Samadi},
  {Goupil}, {Kupka}, \& {Baudin}}]{Kevin06b}
{Belkacem}, K., {Samadi}, R., {Goupil}, M.~J., {Kupka}, F., \& {Baudin}, F.
  2006{\natexlab{b}}, \aap, 460, 183

\bibitem[{{Benomar} {et~al.}(2009){Benomar}, {Baudin}, {Campante},
  {Chaplin}, {Toutain}, \& {et~al}}]{Benomar09b}
{Benomar}, O., {Baudin}, F., {Campante}, T., {et~al.} 2009, \aap, 507, L13, arXiv:0910.3060

\bibitem[{{B\"ohm-Vitense}(1958)}]{Bohm58}
{B\"ohm-Vitense}, E. 1958, Zeitschr. Astrophys., 46, 108

\bibitem[{{Bruntt} {et~al.}(2008){Bruntt}, {De Cat}, \& {Aerts}}]{Bruntt08}
{Bruntt}, H., {De Cat}, P., \& {Aerts}, C. 2008, \aap, 478, 487

\bibitem[{{Chaplin} {et~al.}(2005){Chaplin}, {Houdek}, {Elsworth}, {Gough},
  {Isaak}, \& {New}}]{Chaplin05}
{Chaplin}, W.~J., {Houdek}, G., {Elsworth}, Y., {et~al.} 2005, \mnras, 360, 859

\bibitem[{{Christensen-Dalsgaard} \& {Berthomieu}(1991)}]{JCD91b}
{Christensen-Dalsgaard}, J. \& {Berthomieu}, G. 1991, {Theory of solar
  oscillations} (Solar interior and atmosphere (A92-36201 14-92).~Tucson, AZ,
  University of Arizona Press, 1991), 401--478

\bibitem[{{Freytag} {et~al.}(2002){Freytag}, {Steffen}, \& {Dorch}}]{Freytag02}
{Freytag}, B., {Steffen}, M., \& {Dorch}, B. 2002, Astronomische Nachrichten,
  323, 213

\bibitem[{{Gillon} \& {Magain}(2006)}]{Gillon06}
{Gillon}, M. \& {Magain}, P. 2006, \aap, 448, 341

\bibitem[{{Grevesse} \& {Noels}(1993)}]{GN93}
{Grevesse}, N. \& {Noels}, A. 1993, in Origin and Evolution of the Elements,
  ed. N.~{Prantzos}, E.~{Vangioni-Flam}, \& M.~{Cass\'e} (Cambridge University
  Press), 15

\bibitem[{{Gustafsson} {et~al.}(2008){Gustafsson}, {Edvardsson}, {Eriksson},
  {J{\o}rgensen}, {Nordlund}, \& {Plez}}]{Gustafsson08}
{Gustafsson}, B., {Edvardsson}, B., {Eriksson}, K., {et~al.} 2008, \aap, 486,
  951

\bibitem[{{Houdek}(2006)}]{Houdek06}
{Houdek}, G. 2006, in ESA Special Publication, Vol. 624, Proceedings of SOHO
  18/GONG 2006/HELAS I, Beyond the spherical Sun, Published on CDROM, p. 28.1

\bibitem[{{Houdek} {et~al.}(1999){Houdek}, {Balmforth},
  {Christensen-Dalsgaard}, \& {Gough}}]{Houdek99}
{Houdek}, G., {Balmforth}, N.~J., {Christensen-Dalsgaard}, J., \& {Gough},
  D.~O. 1999, \aap, 351, 582

\bibitem[{{Jacoutot} {et~al.}(2008){Jacoutot}, {Kosovichev}, {Wray}, \&
  {Mansour}}]{Jacoutot08a}
{Jacoutot}, L., {Kosovichev}, A.~G., {Wray}, A.~A., \& {Mansour}, N.~N. 2008,
  \apj, 682, 1386

\bibitem[{{Ludwig}(1992)}]{Ludwig92}
{Ludwig}, H.-G. 1992, PhD thesis, University of Kiel

\bibitem[{{Mosser} {et~al.}(2005){Mosser}, {Bouchy}, {Catala}, {Michel},
  {Samadi}, {Th{\' e}venin}, {Eggenberger}, {Sosnowska}, {Moutou}, \&
  {Baglin}}]{Mosser05}
{Mosser}, B., {Bouchy}, F., {Catala}, C., {et~al.} 2005, \aap, 431, L13

\bibitem[{{Neuforge-Verheecke} \& {Magain}(1997)}]{Neuforge-Verheecke97}
{Neuforge-Verheecke}, C. \& {Magain}, P. 1997, \aap, 328, 261

\bibitem[{{Nordlund}(1982)}]{Nordlund82}
{Nordlund}, A. 1982, \aap, 107, 1

\bibitem[{{Nordlund} \& {Stein}(2001)}]{Stein01I}
{Nordlund}, {\AA}. \& {Stein}, R.~F. 2001, \apj, 546, 576

\bibitem[{{Samadi} {et~al.}(2008{\natexlab{a}}){Samadi}, {Belkacem}, {Goupil},
  {Dupret}, \& {Kupka}}]{Samadi08}
{Samadi}, R., {Belkacem}, K., {Goupil}, M.~J., {Dupret}, M.-A., \& {Kupka}, F.
  2008{\natexlab{a}}, \aap, 489, 291

\bibitem[{{Samadi} {et~al.}(2008{\natexlab{b}}){Samadi}, {Belkacem}, {Goupil},
  {Ludwig}, \& {Dupret}}]{Samadi08b}
{Samadi}, R., {Belkacem}, K., {Goupil}, M.-J., {Ludwig}, H.-G., \& {Dupret},
  M.-A. 2008{\natexlab{b}}, Communications in Asteroseismology, 157, 130

\bibitem[{{Samadi} {et~al.}(2007){Samadi}, {Georgobiani}, {Trampedach},
  {Goupil}, {Stein}, \& {Nordlund}}]{Samadi07a}
{Samadi}, R., {Georgobiani}, D., {Trampedach}, R., {et~al.} 2007, \aap, 463,
  297

\bibitem[{{Samadi} \& {Goupil}(2001)}]{Samadi00I}
{Samadi}, R. \& {Goupil}, M.~. 2001, \aap, 370, 136

\bibitem[{{Samadi} {et~al.}(2006){Samadi}, {Kupka}, {Goupil}, {Lebreton}, \&
  {van't Veer-Menneret}}]{Samadi05b}
{Samadi}, R., {Kupka}, F., {Goupil}, M.~J., {Lebreton}, Y., \& {van't
  Veer-Menneret}, C. 2006, \aap, 445, 233

\bibitem[{{Samadi} {et~al.}(2009){Samadi}, {Ludwig}, {Belkacem}, {Goupil},
  {Benomar}, {Mosser}, {Dupret}, {Baudin}, {Appourchaux}, \&
  {Michel}}]{Samadi09b}
{Samadi}, R., {Ludwig}, H.-G., {Belkacem}, K., {et~al.} 2009, \aap, in press, arXiv:0910.4037 (Paper~II)

\bibitem[{{Samadi} {et~al.}(2003{\natexlab{a}}){Samadi}, {Nordlund}, {Stein},
  {Goupil}, \& {Roxburgh}}]{Samadi02II}
{Samadi}, R., {Nordlund}, {\AA}., {Stein}, R.~F., {Goupil}, M.~J., \&
  {Roxburgh}, I. 2003{\natexlab{a}}, \aap, 404, 1129

\bibitem[{{Samadi} {et~al.}(2003{\natexlab{b}}){Samadi}, {Nordlund}, {Stein},
  {Goupil}, \& {Roxburgh}}]{Samadi02I}
{Samadi}, R., {Nordlund}, {\AA}., {Stein}, R.~F., {Goupil}, M.~J., \&
  {Roxburgh}, I. 2003{\natexlab{b}}, \aap, 403, 303

\bibitem[{{Smagorinsky}(1963)}]{Smagorinsky63}
{Smagorinsky}, J. 1963, Monthly Weather Review, 91, 99

\bibitem[{{Solano} {et~al.}(2005){Solano}, {Catala}, {Garrido}, {Poretti},
  {Janot-Pacheco}, {Guti{\'e}rrez}, {Gonz{\'a}lez}, {Mantegazza}, {Neiner},
  {Fremat}, {Charpinet}, {Weiss}, {Amado}, {Rainer}, {Tsymbal}, {Lyashko},
  {Ballereau}, {Bouret}, {Hua}, {Katz}, {Ligni{\`e}res}, {L{\"u}ftinger},
  {Mittermayer}, {Nesvacil}, {Soubiran}, {van't Veer-Menneret}, {Goupil},
  {Costa}, {Rolland}, {Antonello}, {Bossi}, {Buzzoni}, {Rodrigo}, {Aerts},
  {Butler}, {Guenther}, \& {Hatzes}}]{Solano05}
{Solano}, E., {Catala}, C., {Garrido}, R., {et~al.} 2005, \aj, 129, 547

\bibitem[{{Stein} {et~al.}(2004){Stein}, {Georgobiani}, {Trampedach}, {Ludwig},
  \& {Nordlund}}]{Stein04}
{Stein}, R., {Georgobiani}, D., {Trampedach}, R., {Ludwig}, H.-G., \&
  {Nordlund}, {\AA}. 2004, \solphys, 220, 229

\bibitem[{{Stein}(1967)}]{Stein67}
{Stein}, R.~F. 1967, Solar Physics, 2, 385

\bibitem[{{Trampedach}(1997)}]{Trampedach97}
{Trampedach}, R. 1997, Master's thesis, Master's thesis, Aarhus University
  (1997)

\bibitem[{{V{\"o}gler} {et~al.}(2004){V{\"o}gler}, {Bruls}, \&
  {Sch{\"u}ssler}}]{Voegler04}
{V{\"o}gler}, A., {Bruls}, J.~H.~M.~J., \& {Sch{\"u}ssler}, M. 2004, \aap, 421,
  741

\bibitem[{{Wedemeyer} {et~al.}(2004){Wedemeyer}, {Freytag}, {Steffen},
  {Ludwig}, \& {Holweger}}]{Wedemeyer04}
{Wedemeyer}, S., {Freytag}, B., {Steffen}, M., {Ludwig}, H.-G., \& {Holweger},
  H. 2004, \aap, 414, 1121

\end{thebibliography}
%\input{hd49933.v3.final.bbl}

\Online

%************************************
\appendix
\section{Theoretical calculation of the mode excitation rates for HD~49933}

The mode excitation rate ${\cal P}$ is inversely proportional to the mode
mass ${\cal M}$ (see Eqs.~(\ref{modemass}), (\ref{inertia}) and
(\ref{pow})). 
This is why we can derive ${\cal M}$ and ${\cal M} \, {\cal P}$ separately in order to derive ${\cal P}$ for  HD~49933.
%Then, in order to derive ${\cal P}$ for  HD~49933, we
%can derive separately ${\cal M}$ and ${\cal M} \, {\cal P}$. 

\subsection{Derivation of ${\cal M} \, {\cal P}$}
As pointed out in Sect.~\ref{metal_effect_interpretation}, the kinetic flux $F_{\rm kin} =
{\bar \rho} \, {\tilde u}^3$ is almost unchanged between 
S1 and S0 because both 3D models have the same $T_{\rm eff}$.
This has  also to  be the case for HD~49933 (same $T_{\rm eff}$ and
same $\log~g$ than S0 and S1). Therefore, the
calculation of ${\cal M} \, {\cal P}$ for HD~49933 relies only on the
evaluation of the values reached --~ at a fixed mode frequency ~--  by the 
source functions ${\cal S}_R $ and ${\cal S}_S $.

As seen in Sect.~\ref{metal_effect_interpretation}, $\omega_0 = k_0 \, \tilde{u}$  controls the width
  of $\chi_k$ in a way that the source functions  ${\cal S}_R(\omega_{\rm osc}) $ and
  ${\cal S}_S(\omega_{\rm osc}) $ can be seen as functions of the
  dimensionless ratio   $ \omega_{\rm osc}/\omega_0$. 
The variation of $E$ with $k$ as well as the
variation of $\chi_k$ with $\omega/\omega_0$ are  shown to be similar in
the two 3D simulations.  Furthermore,   S0 and S1 have approximately
the same characteristic length
$\Lambda$ and hence approximately the same $k_0 \equiv 2 \pi / \Lambda$. 
Therefore, the source function ${\cal S}_R$ 
 (resp. ${\cal S}_S $) associated with S0 only differs from that of S1
by the characteristic velocity ${\tilde u}$.
This must then also be the case for HD~49933. 
Further, in order to evaluate the source functions in the case of HD
~49933, we only need to know the factor $\gamma$ by which ${\tilde u}$ is modified in HD
~49933 with respect to S1 or S0. According to Eq.~(\ref{SR})
(resp. Eq.~(\ref{SS})),  multipling  ${\tilde u}$ by  $\gamma$ is
equivalent to replace ${\cal S}_R(\omega_{\rm osc})$ (reps. ${\cal
  S}_S(\omega_{\rm osc})$) by  $\gamma \, {\cal S}_R(\omega_{\rm osc}/\gamma)$
(resp. $\gamma \, {\cal S}_s(\omega_{\rm osc}/\gamma)$). 

Since the kinetic flux $F_{\rm kin}$  in HD~49933 must be the same for
S0 or S1, the characteristic velocity ${\tilde u}$ can be derived  for 
HD~49933 according to ${\tilde u}_* (T) = {\tilde u}_1 \, \gamma_*$
with $\gamma_* (T) \equiv ( {\bar \rho}_1 / {\bar \rho}_* )^{1/3} $
where ${\bar \rho}_1(T)$ is the mean density stratification of S1, ${\tilde u}_1(T)$ the
characteritic velocity of S1 and ${\bar \rho}_*$ the mean density of
HD~49933. 
Once $\gamma_* $ and then ${\tilde u}_*$ are derived for HD~49933, we then compute 
the source functions associated with HD~49933.
 Finally, we compute ${\cal M} \,{\cal P}$ by 
 keeping  $F_{\rm kin}$ constant. We now turn to the derivation of the factor $\gamma_*$.

\subsection{Derivation of ${\gamma}_*$}

To derive $\gamma_*$ at a given $T$, we need to know
how the mean density ${\bar \rho}$ varies with the metal abundance $Z$. 
In order to this we consider five   ``standard'' 1D models with five different values of
$Z$. These 1D models are built  using the same physics as described in Sect.~\ref{1Dmodels}.
Two of these models have the same abundance as S0 and S1.
All of the  1D models have approximately the same gravity ($\log g
\simeq 4.25$) and the same effective temperature ($T_{\rm eff} \simeq
6730$~K).

The set of 1D models shows that --~ at any given temperature within the
excitation region ~-- ${\bar \rho}$ varies with $Z$ rather linearly. 
In order to derive  ${\bar
  \rho}$ for HD~49933, we apply --~ at fixed $T$ and between S0 and S1
~-- a linear interpolation of  ${\bar   \rho}(T)$ with respect to $Z$.

\subsection{Derivation of ${\cal M}$}
As shown in Sect.~\ref{metal_effect_interpretation} above in the frequency domain where modes are
detected in HD~49933, ${\cal M}$ does not change significantly
between S0 and S1.
 This suggests that the mode masses associated with
a patched 1D model with the metal abundance expected for HD~49933
would be very similar to those associated with S0 or S1.
Consequently we will assume for the case of HD~49933 the same mode masses
as those associated with S1, since this 3D model has a $Z$ abundance
closer to that of HD~49933.

\subsection{Derivation of ${\cal P}$}

Before deriving ${\cal P}$ for HD~49933, we check that, from S0 and
the knowledge of ${\cal P}_0$, we can approximately reproduce
${\cal P}_1$, the mode excitation rates,  associated with S1 following the procedure described above.
Let  $\gamma_1  \equiv \left ( {\bar  \rho_0} / {\bar  \rho_1}
\right) ^{1/3}$. As seen
in Fig.~\ref{figu0}, when we multiply ${\tilde u}_0$ by $\gamma_1(T)$ we
matche ${\tilde u}_1$.  Then, using $\gamma_1(T)$ and following the
procedure described above, we derive  ${\cal P}_{01}$, the mode excitation rates
associated with S1 but derived from S0. The result is shown in Fig.~\ref{pow2}.  ${\cal P}_{01}$ matches
 ${\cal P}_1$ rather well. 
However, there are  differences remaining in particular in the frequency domain $\nu$=1.2-1.5~mHz. %They are due to the position of the nodes of the eigenfunctions. Indeed, these nodes are not exactly placed at the same position for S0 and for S1.
Nevertheless, the differences between ${\cal P}_{01}$ and $\cal P_1$  are in any case not significant compared to the accuracy at which the mode amplitudes are measured with the CoRoT data (see Paper II). 
 This validates the procedure, at least at the level of the current seismic precisions. 
%However, this maximum peaks at lower
%frequency, that is at $\nu_{\rm max} \simeq 1.4$~mHz instead at
%$\nu_{\rm max} \simeq 1.7$~mHz for ${\cal P}_1$. This is likely because we use
%eigenfunctions that are not representative for S1.
\fig{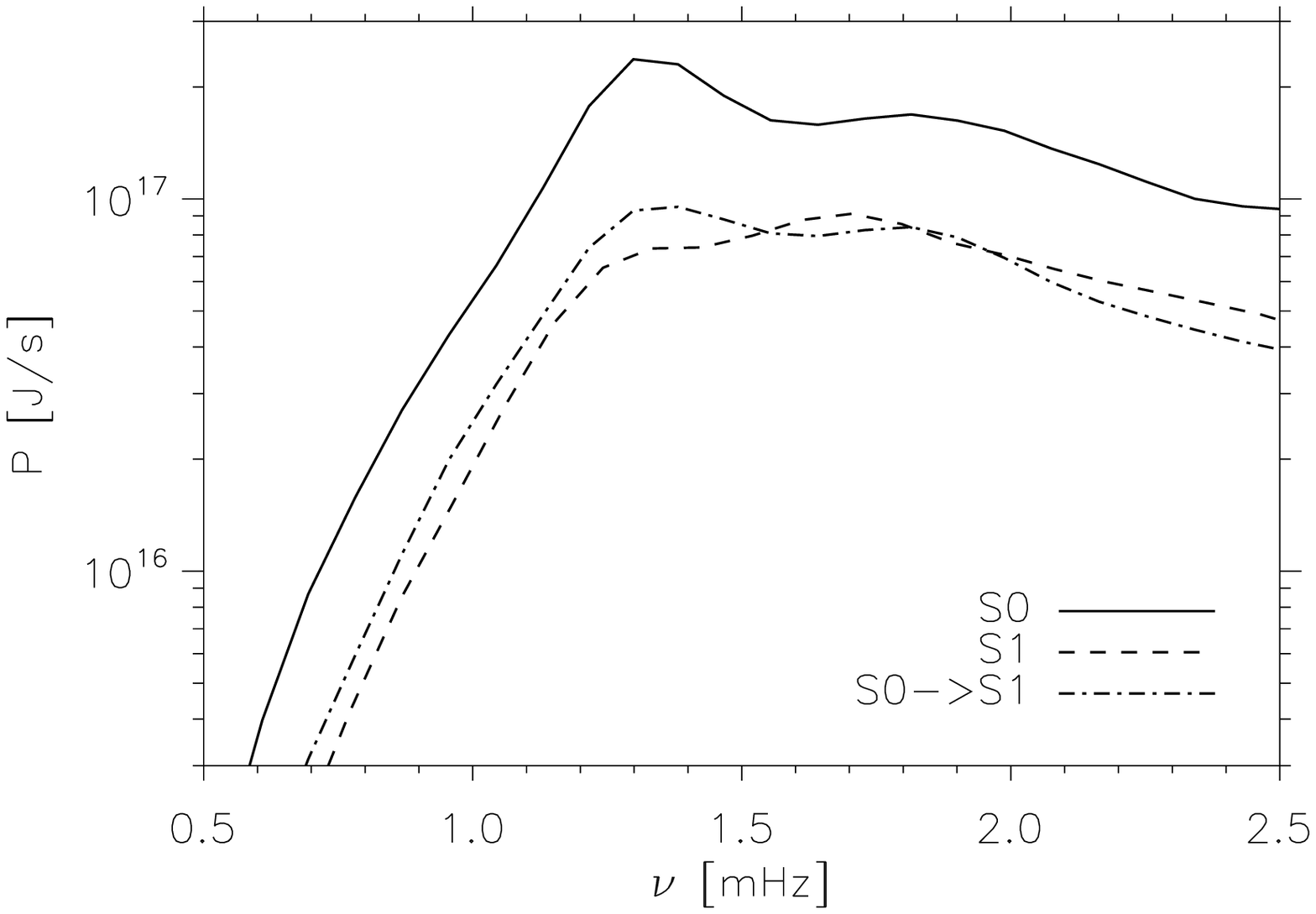}{Mode excitation rates $\mathcal P$ as a function
of the mode frequency $\nu$. 
The thin dot-dashed line corresponds to ${\cal P}_{01}$, the mode excitation rates
derived for S1 from S0 (see Appendix~A4). The other  lines have the same
meaning as in Fig.~\ref{pow}.
}{pow2}

Since the metal abundance Z of HD~49933 is closer to that of S1 than that of S0, we
derive the mode excitation rates ${\cal P}$  associated with
HD~49933 from S1 following the procedure detailled above. 
The result is shown in Fig.~\ref{pow}. 
As expected, the
mode excitation rates ${\cal P}$  associated with HD~49933 lie 
between those of S0 and S1, while remaining closer to S1 than to S0.  
Note that the differences between ${\cal P}_1$ and the excitation
rates derived for HD 49933 ($\cal P$) are of the same order as
the differences seen locally between  ${\cal P}_1$ and ${\cal P}_{01}$. 
These differences remain small compared to the current seismic precisions. 
On the other hand the differences between $\cal P$ and  ${\cal P}_0$  are significant and
have an important impact on the mode amplitudes (see Paper~II).

\end{document}